\documentclass[epjST]{svjour}
\usepackage{graphics}
\usepackage{framed,xcolor}
\usepackage[english]{babel}
\usepackage{mdframed,xcolor}
\usepackage{hyperref}
\hypersetup{
  colorlinks = true,
  urlcolor = blue}
\begin{document}
\title{Introduction: The FuturICT Knowledge Accelerator Towards a More Resilient and Sustainable Future}

\author{Dirk Helbing}
\institute{ETH Zurich, Swiss Federal Institute of Technology, Switzerland}
\abstract{
The FuturICT project is a response to the European Flagship Call in the Area of Future and Emerging Technologies, which is planning to spend 1 billion EUR on each of two flagship projects over a period of 10 years. FuturICT seeks to create an open, global but decentralized, democratically controlled information platform that will use online data and real-time measurements together with novel theoretical models and experimental methods to achieve a paradigm shift in our understanding of today's strongly interdependent and complex world and make our techno-socio-economic systems more flexible, adaptive, resilient, sustainable, and livable through a participatory approach.
} 
\maketitle
\section*{Introduction}
\label{intro}

Today, we have a better understanding of our universe than about the way our society and economy evolve. We have developed technologies that allow ubiquitous communication and instant information access, but we do not understand how this changes our society. Challenges like the financial crisis, the Arab spring revolutions, global flu pandemics, terrorist networks, and cybercrime are all manifestations of our ever more highly interconnected world. They also demonstrate the gaps in our present understanding of techno-socio-economic-environmental systems.
\par
In fact, the pace of global and technological change, in particular in the area of Information and Communication Technologies (ICTs), currently outstrips our capacity to handle them. To understand and manage the dynamics of such systems, we need a new kind of science and novel socially interactive ICTs, fostering transparency, trust, sustainability, resilience, respect for individual rights, and opportunities for participation in political and economic processes.
\par
As Columbia University's president Lee C. Bollinger put it: ``\emph{The forces affecting societies around the world ... are powerful and novel. The spread of global market systems ... are ... reshaping our world ... raising profound questions. These questions call for the kinds of analyses and understandings that academic institutions are uniquely capable of providing….}"
\par
Information and Communication Technologies (ICTs) are increasingly playing a key role for the occurrence, understanding and solution of many problems that our society is facing. Our global ICT system consists of billions of connected components, including not just computers and smartphones, but increasingly cars, homes, and factories. Many devices take autonomous decisions, based on real-world data, an internal representation of the outside world, and expectations regarding the future. In doing so they influence not just their own state, but impact the real world and human behavior. Already today, supercomputers perform most financial transactions in the world.
\par
It is safe to say that the global ICT system is the most complex artifact ever created, rivaling in complexity with the human brain, human society and the global ecosystem. Although humans have built the individual components that compose the system, we are more and more losing the ability to understand the system as a whole and its interaction with society. In fact, ICT systems are themselves increasingly becoming something like `Artificial Social Systems'. However, they are not constructed in a way that ensures beneficial societal outcomes. The result can be breakdowns of coordination and performance as well as `tragedies of the commons', instabilities, conflicts, (cyber-)crime, and (cyber-) war. A deep understanding of social systems will be needed to get these systems right and also to mitigate our societal problems.
\par
Moreover, we have now a global exchange of people, goods, money, information, and ideas, which has created a strongly coupled and strongly interdependent world. This often causes feedback and cascading effects, extreme events, and unwanted side effects. In fact, these systems behave fundamentally different from weakly coupled systems. Multi-component systems can be dynamically complex and hardly controllable. This requires a paradigm shift in our thinking, moving our attention from the properties of the system components to the collective behavior and emergent systemic properties resulting from the interactions of these components.
\par
As the paradigm shift from a geocentric to a heliocentric world view has facilitated modern physics, so will this paradigm shift towards an interaction-based, systemic perspective and a co-evolution of ICT with society open up entirely new solutions to address old and new problems, for example, financial crises, social and political instabilities, global environmental change, organized crime, the quick spreading of new diseases, and the requirement to build smart cities and energy systems.
\par
In fact, there are promising new approaches to manage complexity: While external control of complex systems is hardly possible due to their self-organized dynamics, one can promote a favorable self-organization by modifying the interaction rules and institutional settings. This adaptive, self-regulating approach has been impressively demonstrated for urban traffic light control and a number of other burning problems. It requires real-time sensing, short-term anticipation, and the implementation of suitable interaction rules between connected system elements. The decentralized self-regulatory principle can be scaled up to systems of almost any size and any kind.
\par
All of this calls for a significant shift in the research and educational focus of academic institutions. Specifically, one needs to develop a better, holistic understanding of the global, strongly coupled and interdependent, dynamically complex systems that humans have created. It is required to push complexity science towards practical applicability, to invent a novel data science, to create a new generation of socially interactive, adaptive ICT systems, and to develop entirely new approaches for systemic risk assessment and integrated risk management.
\begin{figure}
\center
\resizebox{0.5\columnwidth}{!}{%
\includegraphics{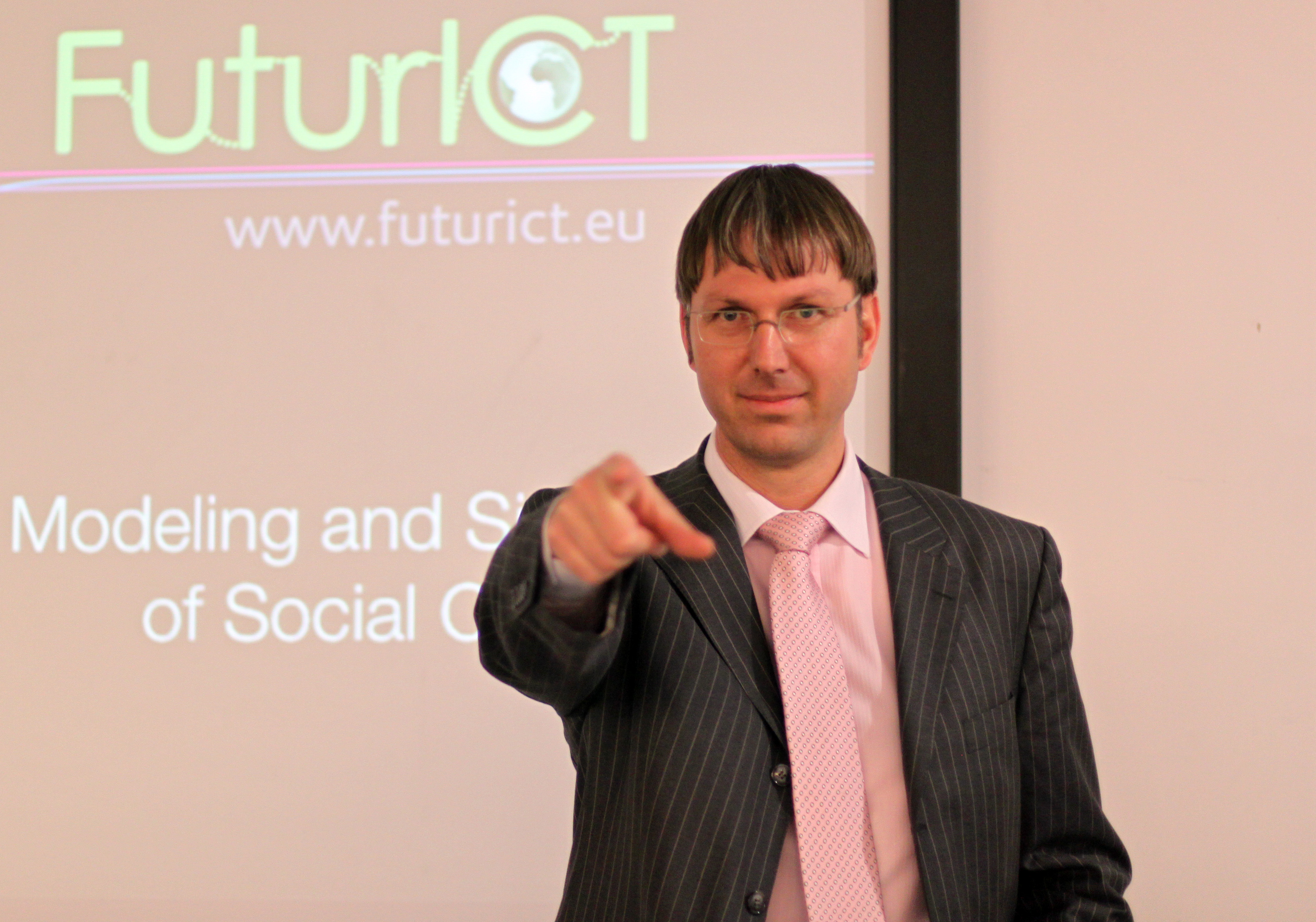}}
\caption{Dirk Helbing, Initiator  and Chair of the Scientific Committee of the FuturICT Flagship project (Photo: Peter Bentley).}
\label{fig:phone}       
\end{figure}
\par
The FuturICT flagship project is the perfect instrument to develop the new science needed to better understand our 21$^{st}$ century challenges, as well as the technology to address and mitigate them. This includes the fundamentals of a `Planetary Nervous System' to enable collective, ICT-based awareness of the state of our world, a `Living Earth Simulator' to explore side effects and opportunities of human decisions, a `Global Participatory Platform' to create opportunities for social, economic and political participation, an `Open Data Platform' to foster the creativity of people and new business opportunities, novel design and operation principles for self-organizing, trustworthy, reliable and safe information exchange, and `Ethical ICT' supporting value-oriented and responsible (inter)action.
\par
In fact, Europe could be leading the upcoming age of social and socio-inspired innovation, which comes with enormous societal and economic potentials.
\par
This special journal issue collects a number of position papers that try to formulate first steps towards the new science and technology required to manage our future more successfully. These papers are no scientific progress reports, but reflections about what could or should be done scientifically in the years to come. They reflect ideas that have been inspired by a number of FuturICT-related workshops. I feel that these workshops have unleashed a great creativity and led us to many interesting new questions that we did not think about in the beginning.
\par
It must also be pointed out that, since the time when these position papers were finalized, the project vision has again moved ahead quite a lot. The FuturICT community is now waiting for support of the envisaged research activities. Let us hope, it will all work out nicely. We are all very excited about the multi-disciplinary and global community that has formed, currently spanning Europe, America, Asia, and Australia, and waiting for Africa to join. I am sure, together we can think our way forward towards more resilient and sustainable societies and new ways of promoting human well-being.

\section*{Appendix: New Ways to Promote Sustainability and Social Well-Being in a Complex, Strongly Interdependent World }

\subsection*{Speech delivered at the ``Out of the Box Conference with World Thinkers", Maribor, May 16, 2012.}
Since the origin of human civilization, there is a continuous struggle between chaos and order. While chaos may stimulate creativity and innovation, order is needed to coordinate human action to create synergy effects, more efficiency, and common goods such as our transportation infrastructures, universities, schools, and theaters, institutions (like parliaments and courts), but also language and culture.
\par
According to Hobbes, civilization started with everyone fighting against everybody else ("homo hominis lupus"). Even today, civilization is highly vulnerable, as the outbreak of war in former Yugoslavia has shown, but also the situation in many countries today, particularly after natural disasters.
\par
On the one hand, we are struggling with conflicts, which are the result of suppression and lack of participation, and of sanctioning people who are not part of the mainstream culture. On the other hand, we are suffering of many social dilemmas, such as the exploitation and destruction of our environment, global warming, overfishing, tax evasion, exploitation of social benefit systems, and other tragedies of the commons.
\par
In order to mitigate problems like these, we need to learn how to understand and manage complexity in our techno-socio-economic-environmental systems. For this, we need to think out of the box, because complex systems work differently from what our intuition suggests.
\par
Complex systems are often hard to predict and hard to manage. And they behave in surprising ways. Their behavior is not well understood from the properties of their components. It's rather the interactions between these components, which we need to focus on, because they are the basis of the self-organization of complex systems and of new, so-called emergent properties, which cannot be understood from the component properties. Society is more than the sum of its parts.
\par
The change of perspective from a component-oriented to an interaction-oriented view may be AS hard and revolutionary as the transition from the geo-centric to the helio-centric worldview, which made modern physics possible, and many of the benefits that came with it. This new systemic view will enable us to find new solutions to old problems such as social conflicts and tragedies of the commons.
\par
Let me stress that economic value generation would not be possible without many kinds of social capital, such as trust, solidarity, social values, norms, and culture. While absolutely crucial for social well-being and the functioning of society, social capital is largely invisible, and hardly understood. We know, however, that it is the result of our social network interactions. But social capital may be damaged or exploited, as the environment has been damaged and exploited. Hence, we need to learn how to value and protect social capital. The evaporation of trust during the financial crisis may serve as a warning example. It caused the evaporation of thousands of billions of dollars in the stock markets and elsewhere.
\par
In order to learn how to protect social capital, we need, first of all, a global systems science, which allows us to gain a holistic understanding of our world, of systemic risks, and how integrated systems design can create more resilient and sustainable systems. Changing interactions in the system can mitigate problems, since the current kinds of interactions are causing them. In fact, unstable interactions cause problems such as tragedies of the commons, segregation, conflict, revolutions, wars, and cascading effects causing extreme events and disasters. Moreover, our global markets are constructed in a way that ethical behavior tends to have a competitive disadvantage that's why ethical behavior has so hard times to survive and spread.
\par
We may understand such systemic instabilities as situations, in which a situation gets out of control even, if everyone is trying to do his/her best. Take dense, but continuous traffic flow on a circular road as an example. Sooner or later, the traffic flow will break down, creating a traffic jam, although everyone tries to avoid it. In other words, systemic instabilities makes systems largely uncontrollable. But to some extent, this can be changed!
\par
Societies have found many ways to create social order. The most archaic one is the creation of families and tribes, which builds on the mechanism of genetic inheritance. Neighborhood interactions (even with strangers) and direct reciprocity based on repeated interactions (I help you and you help me) is a more modern mechanism to promote social cooperation. Furthermore, humans have created sanctioning institutions (including the police). And last but not least, cooperation and social order may be promoted by reputation mechanisms. To counteract global destabilization, we are currently seeing a growth in sanctioning efforts, but reputations systems are also quickly spreading, e.g. in the internet.
\par
To stabilize our ever more complex systems, we will have to integrate decentralized management elements into our management approaches, utilizing the principle of self-organization. New information and communication technologies make it possible to overcome barriers to social, economic and political participation. They could also support fair ways of sharing. Again, the interaction rules are crucial here. If someone who cuts a cake can take the first piece, it will often be the biggest one. If the person, who cuts the cake takes the last piece, he/she will be very careful to cut it in a fair way, with all pieces having an equal size.
New information and communication technologies can also promote an inter-cultural understanding (via an intercultural translator), facilitate social money (with a memory and reputation), or value-sensitive action. They can furthermore promote accountability and awareness.
\par
Awareness helps to avoid many mistakes one would otherwise make. The FuturICT project plans to create a Planetary Nervous System, which will support such awareness of the state of the world, including the value of social capital. FuturICT's Living Earth Simulator will help to anticipate possible scenarios (as we have done it by mental simulation in the past for much simpler situations). This can warn us of systemic risks, but also point us to new opportunities. And finally, FuturICT's Global Participatory Platform will make these new instruments, which serve to gain insights into our complex world, accessible to everyone.
\par
Remember, most people do want to appear attractive and beautiful. If we manage to create awareness of the implications of their decisions and actions, they will change their behavior. I am deeply convinced that we can create a better world - if we create suitable instruments, gain better insights, and do it together.

\section*{Acknowledgments}
The publication of this work was partially supported by the European
Union's Seventh Framework Programme (FP7/2007-2013) under grant agreement no.284709, a Coordination and Support Action in the Information
and Communication Technologies activity area (`FuturICT' FET Flagship Pilot Project).

\end{document}